\documentstyle[prl,aps,floats,twocolumn]{revtex}
\input psfig.tex
\begin{document}
\draft
\twocolumn[\hsize\textwidth\columnwidth\hsize\csname @twocolumnfalse\endcsname
\preprint{DAMTP-96-44, astro-ph/9604172}
\title{Causality and  the Doppler Peaks}
\author{Neil Turok}
\address{DAMTP, Silver St,\\
 Cambridge, CB3 9EW,  U.K.\\
Email: N.G.Turok@damtp.cam.ac.uk\\
}
\date{23/7/96}
\maketitle

\begin{abstract}
A considerable experimental effort is underway to 
detect the `Doppler peaks' in the angular power spectrum of 
the cosmic microwave anisotropy. 
These peaks offer unique 
information about structure formation in the universe.
One  key issue is whether structure could have
formed by the action of causal physics
within the standard hot big bang, 
or whether a prior period of
inflation was required. Recently there has been some discussion of whether
causal sources could reproduce the pattern of Doppler peaks 
produced by the standard adiabatic theory.
This paper gives a rigorous definition of
causality, and a causal decomposition of a general source.
I present an example of a very simple
causal source which  mimics the 
standard adiabatic theory, accurately reproducing the behaviour of
the local intrinsic temperature perturbations.

\end{abstract}
\hspace{0.2in}
]

Existing theories of cosmic structure formation are of two types.
In the first,
the hot big bang is assumed to have started 
out smooth. Structure then forms as
the result of a symmetry breaking phase transition and
phase ordering. 
In the second, 
an epoch of inflation {\it prior} to the hot big bang is invoked. 
Whilst both mechanisms
are causal, causality imposes a much stronger constraint in the former case
(Figure 1),
because the initial conditions for the perturbation variables are established
on a Cauchy surface $\Sigma$ {\it within} 
 the hot big bang (Figure 1).
The causal nature of the Einstein-matter field equations 
then implies the 
vanishing of 
all correlations between all local perturbation 
variables at spacetime points whose
backward light cones fail to intersect on $\Sigma$.
In the inflationary case, by construction
the relevant surface $\Sigma$ lies so far before $\tau=0$ that
there is no useful constraint.

\begin{figure}[htbp]
\centerline{\psfig{file=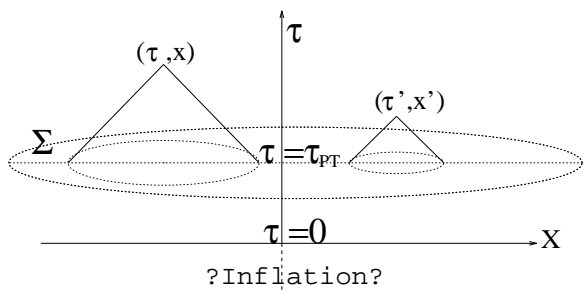,width=3in}}
\caption{ Causal constraint on theories of structure formation 
where the standard hot big bang starts out homogeneous. The vertical 
axis shows conformal time $\tau$, with $\tau=0$ corresponding to 
the initial singularity (in the absence of inflation).
Correlations between any local variables  at any two
spacetime points vanish if their backward light cones fail to intersect
on $\Sigma$, the spacelike hypersurface $\tau=\tau_{PT}$
just prior to the phase transition.
In inflationary theories, there is no singularity at $\tau=0$, instead
there is a preceding epoch of inflation during 
which longer range correlations are established.
}
\label{fig:f1}
\end{figure}

Could observations distinguish the causally constrained theories
from inflationary ones?
The cosmic microwave anisotropy is the best hope of
a direct probe, giving 
us a picture of the universe on the surface of last scattering.
This surface 
cuts through many regions which were `causally disconnected' 
(quotes indicate a standard hot big bang definition) at that
time. If the only 
contributions to the microwave anisotropy were local effects, like
temperature and velocity perturbations in the photon-baryon fluid, 
one could check whether `superhorizon'
perturbations 
were present by 
measuring the autocorrelation function of  the anisotropy map.
If this was  consistent with
zero beyond some angular scale (twice that 
subtended by the `causal 
horizon' at last scattering, of order $2^o$ with standard recombination),
one could conclude that 
the perturbations were indeed causally
constrained.

The complication that spoils this test is that a significant
component 
of the microwave 
anisotropy is generated after last scattering, 
by the integrated effect of time dependent 
gravitational potentials along photon paths. This is after all
how 
cosmic defects produce a scale invariant spectrum of 
microwave anisotropies on very large angular scales 
(consistent with the COBE results) even though these theories {\it are}
causally constrained.

Nevertheless, the local contributions to the microwave anisotropy
do have a signature distinguishing them
from the foreground due to the integrated effect.
This is the presence of 
`Doppler' peaks in the angular power spectrum,
caused by phase-coherent 
oscillations in the photon-baryon fluid prior to recombination.
The location of these 
peaks is mainly determined by the temperature
perturbations in the photon-baryon fluid, a completely local effect.
This Letter will address
the question of whether the peak locations
 can be used as a discriminator
between inflationary and non-inflationary 
theories of structure formation.

Crittenden and I suggested a connection between causality and 
peak location \cite{ct}
following an analysis of the cosmic texture theory, in which the
Doppler peaks are phase-shifted  relative to those in standard 
inflation, the biggest peak occurring at
higher multipole $l$ 
(smaller angular scales) than in the standard inflationary theory.
Albrecht, Magueijo and collaborators
\cite{albrecht} raised the important issue of decoherence, 
and gave a detailed 
discussion of the behaviour of the Doppler 
peaks for different models of causal sources, in particular those
motivated by the study of Robinson and Wandelt \cite{rw}.
Most recently Hu and White \cite{hw} have addressed these issues,
claiming that the pattern of Doppler peaks predicted 
by the simplest inflation model is `essentially unique and its confirmation 
would have deep implications for the causal structure of the early universe'. 
In this letter I develop
a formalism for dealing with decoherent but causal sources.
I exhibit 
a causal source which closely 
mimics the inflationary 
pattern 
of temperature
perturbations in the photon-baryon fluid, and the corresponding
contribution to the microwave anisotropy.

Structure formation within the standard hot big bang requires the 
presence of a source term in the Einstein equations in addition 
to the usual metric and matter variables. Cosmic string and texture
each provide an example of such a sources. The perturbations are most simply
dealt with in the fluid approximation, which is reasonable for
our purposes. In the synchronous gauge, the relevant equations read \cite{sv}
\begin{equation}
\ddot{\delta_C} + {\dot{a}\over a} \dot{\delta_C} =
 4 \pi G 
\sum_N (1+3 c_N^2) \rho_N a^2 \delta_N + {\cal S}, 
\label{eq:c1}
\end{equation}
\begin{equation}
\ddot{\delta_R} + {\dot{a} \over a}(1-3 c_S^2) \dot{\delta_R} =
c_S^2 \nabla^2 \delta_R + {4\over 3} \ddot{\delta_C} + 
{4\over 3} {\dot{a} \over a}(1-3 c_S^2) \dot{\delta_C}
\label{eq:c2}
\end{equation}
Dots denote derivatives with respect to $\tau$, 
$a(\tau )$ is the scale factor, $\delta_C$ and $\delta_R$ are the 
contrast in dark matter and radiation densities, and $c_S$ is the speed of 
sound. The sum over $N$ includes dark matter, 
the photon-baryon fluid and neutrinos. I assume
the `canonical' parameter values  $\Omega_{CDM}=0.95$, $\Omega_{B}=0.05$,
and $h=0.5$.
The fluctuating part of the
external source is taken to have stress energy tensor $\Theta_{\mu \nu}$,
and ${\cal S} = 4\pi G(\Theta_{00}+ \Theta_{ii})$.
The initial conditions to be used with 
(\ref{eq:c1}) and (\ref{eq:c2}) are the vanishing of the pseudoenergy
$\tau_{00}= \Theta_{00} + \sum_N \rho_N a^2
 \delta_N +(\dot a / a) \dot{\delta_C} /
(4 \pi G)$, and adiabaticity, $\delta_R=\delta_\nu={4\over 3}\delta_C$.
This corresponds to starting with a universe in which the energy density
and space curvature are uniform.

It is a good approximation to treat the radiation as 
a perfect fluid, obeying
the equations (\ref{eq:c1}) and (\ref{eq:c2})
from initial conditions set up deep in the radiation era, up to
recombination. The intrinsic temperature perturbation 
from a Fourier mode ${\bf k}$ 
is then given by $(\delta T/T)_i({\bf k})
 = {1\over 4} \delta_R({\bf k})$, and its contribution to the 
angular power spectrum of anisotropies is 
given by $C_l \propto \int k^2 dk 
(\delta T/T)_i^2(k) j_l(k\tau_0)^2$ with $\tau_0$ the conformal time today
\cite{efstath}. This dominates the anisotropy pattern on small angular scales.

Equations (\ref{eq:c1}) and (\ref{eq:c2}) are linear, and it follows that
all correlations between local observables are completely 
determined by the unequal time correlation function of
the source stress energy tensor.
In particular, for ${\cal S}$ the causality constraint reads 
\begin{equation}
\xi(r,\tau,\tau')= 
<{\cal S}(r,\tau) {\cal S} (0,\tau')> = 0 \qquad r> \tau+\tau'
\label{eq:c3}
\end{equation}
The sharp edge on $\xi$ leads to 
oscillations in its three dimensional 
Fourier transform $\tilde{\xi} (k,\tau,\tau')$ at large $k$.
Integration by parts produces 
\begin{equation}
\tilde{\xi} (k,\tau,\tau') \sim  [-{R \over k^2}
{\rm cos} kr 
+ {R_{,r} \over k^3}{\rm sin} kr + ... ]^{\tau+\tau'}_0,
\label{eq:c4}
\end{equation}
where $R(r) = r \xi(r)$, 
and $R_{,r}=dR/dr$. (If $\xi \sim r^{-2}$
at small $r$, as it does for strings,
then $\tilde{\xi}$ has an additional $k^{-1}$ term).
The leading term
is not necessarily oscillatory, but there {\it must} 
be oscillatory subleading terms. Most of the  ansatzes 
for $\tilde{\xi}$ 
in the literature do not have this feature and are 
therefore manifestly acausal. They may still be useful as 
approximations, but it is desirable 
to develop a formalism
in which causality is rigorously built in.

The discussion simplifies if we assume scaling
\cite{transn}.
Then dimensional analysis implies that 
\begin{equation}
\tilde{\xi} = <S({\bf k},\tau) S^*({\bf k},\tau')>=
\tau^{-{1\over 2}}  \tau'^{-{1\over2}} X(k\tau,k\tau').
\label{eq:c5}
\end{equation}
Regarded as a matrix 
with indices $\tau$, and $\tau'$, 
$X$ is real and symmetric and can therefore be represented as:
\begin{equation}
X(k\tau,k\tau')= \sum_\alpha P_\alpha 
f_\alpha(k\tau)
f_\alpha(k\tau')
\label{eq:c7}
\end{equation}
with $f_\alpha(k\tau)$ a set of 
orthonormalised
eigenfunctions of $X(k\tau,k\tau')$ regarded as an integral operator,
with corresponding eigenvalues 
$P_\alpha$.
In the terminology of \cite{ct}, $f_\alpha(k\tau)$ is a `master' function.
As in quantum mechanics,
we have a pure, `coherent' state if $P_\alpha$ is nonzero for 
only a single value of $\alpha$, otherwise we have a mixed, `incoherent' state. 
Equation (\ref{eq:c7})
shows that a general source may be represented as an incoherent sum of
coherent sources. The $P_\alpha$'s must be 
positive for all $\alpha$ 
because they are the expectation value of a quantity squared.

This representation is useful because the contribution 
of each individual term in the 
$\alpha$ sum is straightforwardly calculable, by using the source
$\tau^{-{1\over 2}} f_\alpha(k\tau)$ in the linearised Einstein equations.
A bonus is that 
the assumption of scaling allows one to 
infer the correct
initial
conditions for the perturbations. For small $k$, (\ref{eq:c7}) assures us
that, if $\tilde{\xi}(0)$ exists, 
$f(k\tau)$ must tend to a (possibly zero)
constant. Then energy conservation,
$\dot \Theta_{00} + (\dot a /a) (\Theta_{00}
 +\Theta_{ii}) \approx 0$ for $k\tau <<1$, and the assumption of scaling
(by dimensions, $\Theta_{00}\propto \tau^{-{1\over 2}}$)
allow one to unambiguously 
determine the contribution to $\Theta_{00}$ appropriate 
to each $f_\alpha(k\tau)$.

Now let us return to the causality constraint (\ref{eq:c3}). Consider
a single term in the sum over $\alpha$ in (\ref{eq:c7}).
The contribution it makes to 
$<S(r,\tau) S(0,\tau')>$ is  proportional to the convolution
of $f_\alpha(r,\tau)$ with $f_\alpha(r,\tau')$. 
If the $f_\alpha(r,\tau)$ have compact support,
a simple argument \cite{proof} shows that 
$f_\alpha(r,\tau)= 0$ for all $r>\tau$.
This gives a nice geometrical 
picture of how the causality constraint works. 
For each $\alpha$ 
the master function $f_\alpha(r,\tau)$ is the profile of a 
ball of radius $\tau$, and the convolution of $f_\alpha(r,\tau)$ with
$f_\alpha(r,\tau')$  clearly
vanishes if the separation 
of the ball centers is greater than $\tau+\tau'$.

Determining the form of the 
$f_\alpha$ and $P_\alpha$ would be very interesting in 
any particular causal scenario. 
Here however, I want to see whether anything useful can be 
learnt by considering all possible
$f_\alpha$'s and $P_\alpha$'s.
The power spectrum in the general case is
is just a sum of the power spectra for different such $f$'s
with positive coefficients, so if for example
we can show that the $C_l$ for every 
$f$ has positive slope for $l<l_{max}$, it follows
that the total power spectrum will too. In this way we can set a
lower 
limit
on the location of the first Doppler peak. 

A basis for all functions 
$f(r)$ is  provided by the family $r^2 f(r,\tau) = \delta(r-A\tau)$,
with $0<A<1$. 
In Fourier space we have
$f(k,\tau)= {\rm sin} Ak\tau/(Ak\tau)$. In one extreme, with
only short range correlations, $f(k,\tau)$ is
nearly constant. In the other, it has its first zero at $k \tau=\pi$. 
The equal time correlation functions corresponding to
this family of master functions are smooth functions: one find
$\xi(r,\tau,\tau)
\sim 1/(r \tau^3)$ for $r<2A\tau$, $\xi(r)=0$ for $r>2A\tau$.
If   
the master functions $f(r)$
do not change sign for all $r<\tau$ (which is unlikely),
 then any $f(r)$ can be represented
as a sum of the above basis functions with positive coefficients.

We now proceed to solve equations 
(\ref{eq:c1})
and (\ref{eq:c2}) for this family of  
source functions, with  ${\cal S} = f_k(\tau)/\tau^{1\over 2}$.
Each Fourier mode of $\delta_R$ 
starts out small and grows (like $\tau^{3\over2}$).
After horizon crossing
it oscillates as an acoustic wave. 
At the `instant' of last scattering, all modes are caught at 
a particular phase of their oscillations, and those
which are at maximum amplitude produce the Doppler peaks in $C_l$.
Figure 2 shows the time evolution of a single Fourier  mode of $\delta_R$
and $\delta_C$ in the two extreme cases ($A=0$, denoted $O$,
and $A=1$, denoted $X$), and in the standard adiabatic theory with no
source. 
The approximate scale invariance means that
the same graph very roughly represents
$\delta_R(k,\tau_{rec})$
as a function of $k$ at recombination. 
One can translate $k\tau_{rec}$ into
multipole moment $l$ by the approximate relation $l\sim k\tau_0 
\sim 50 k\tau_{rec}$.
Peaks in $\delta_R(k,\tau)^2$ are, through the integral given above, 
translated into peaks in $C_l$.

\begin{figure}[htbp]
\centerline{\psfig{file=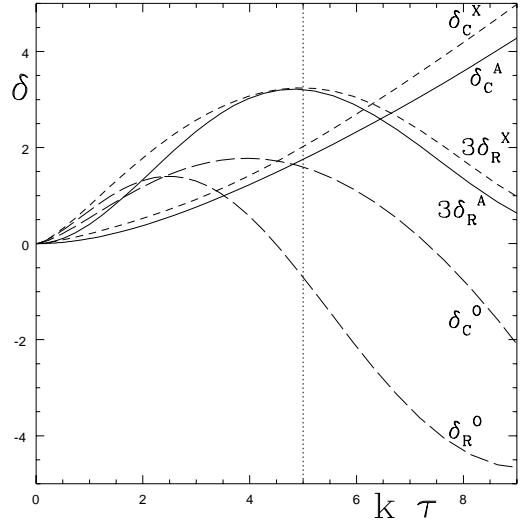,width=3in}}
\caption{ The evolution of perturbation modes for radiation
$\delta_R$ and dark matter $\delta_C$ as a function of conformal time
$\tau$, for fixed $k$.  The value of $k$ chosen has $k\tau_{rec}=5$,
and in the inflationary theory 
the corresponding mode has its first maximum at recombination. 
The superscript $A$ denotes the standard 
inflationary theory, 
$X$ shows the extreme causal model which mimics
the case $A$, and $O$ shows the opposite extreme
causal model which mimics instead
the texture model. 
}
\label{fig:f2}
\end{figure}

In the causal theories, $\delta_C$ is forced to 
start out growing with sign
{\it opposite} to the source ${\cal S}$, because 
the total pseudoenergy $\tau_{00}$ 
must initially be zero. 
As time goes on, ${\cal S}$ starts to drive $\delta_C$.

If ${\cal S}$ always has the same sign, as in the case 
$A <<1$,
$\delta_C$ changes sign as it becomes driven by ${\cal S}$.
The forcing term for the
radiation, $\ddot \delta_C$, then changes
sign around $k\tau=1$, so while $\delta_R$ initially grows
with the opposite sign to ${\cal S}$, it is later driven to the same sign 
as ${\cal S}$. Because the sign 
change in  $\ddot \delta_C$ occurs early
(at $k\tau \sim 1$) the first oscillation in the radiation has 
small amplitude. Because $C_l$ is really an integral over $k$, as
mentioned above, the first Doppler peak, at $l\sim 120$, is smeared
by the contribution of higher $k$, and may be
effectively `hidden'. The main peak is that due to the next oscillation,
the one that is really `driven' by ${\cal S}$. This one occurs at 
$l\sim 380$, compared to the main peak in the standard adiabatic case 
at $l\sim 220$.  Inside the horizon, the radiation 
oscillates sinusoidally,
and  higher peaks occur at  shifts 
$\Delta l \approx  280n$, $n=1,2,3,..$ to the right.
It is interesting that this case ($A<<1$) reproduces the main features
of the texture models presented in \cite{ct} and \cite{durrer}.

Next, consider the case where there {\it is}
a sign change in ${\cal S}$ 
around horizon crossing. 
As before $\delta_C$ and $\delta_R$
start out with the opposite sign to ${\cal S}$, because of compensation. 
But here, if ${\cal S}$ changes sign early enough, $\delta_C$
{\it does not} have to change sign. The radiation forcing 
term  $\ddot \delta_C$ is always positive, and the first peak in $\delta_R$ 
is not small. As can be seen in Figure 2, the extreme case
$A=1$ mimics the standard adiabatic model rather closely.
I have computed the power spectrum $\delta_R(k,\tau)^2$ at recombination in
the $A=1$ theory, for $\Omega_B=0.05,$ $0.1$, and $0.2$ (and $\Omega_{CDM}
=1-\Omega_B$). In all cases
the result is similar
to the analogous standard adiabatic theory, both in the peak location and
in the pattern of peak heights.

It may be useful to visualise  this in terms of the $C_l$ spectra.
Figure 3 contrasts the standard
adiabatic $C_l$'s with those of the texture model \cite{ct}, \cite{durrer}.
The simple family of causal theories I have just discussed
roughly speaking interpolates between these two curves. 
At $A=1$ the first peak is close to that of the 
adiabatic theory, and as
one decreases $A$, the 
peaks move to lower $l$.
The first
peak decreases in amplitude, for the reason discussed 
above, and moves leftward from 
$l\sim 220$ to $l\sim 120$. The second increases in amplitude, and 
moves from $l\sim 500$ to $l \sim 380$.

\begin{figure}[htbp]
\centerline{\psfig{file=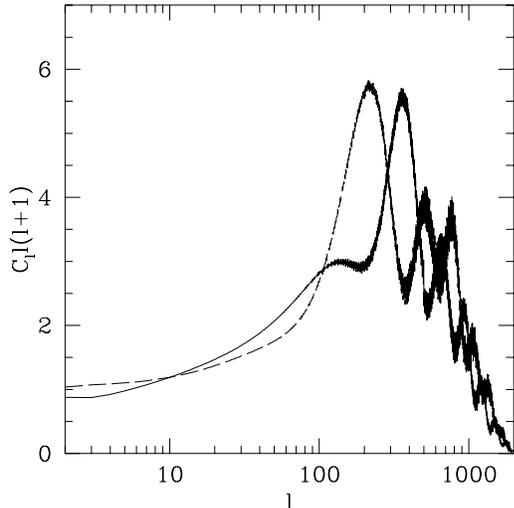,width=2.8in}}
\caption{ The  
anisotropy power spectra for the standard 
inflationary theory (dashed line) 
and the texture theory (solid line).
The family of sources studied here produces peak locations which
approximately (at the ten per cent level) interpolate between
these two cases.
As the parameter $A$ is dialed 
from 1 to 0, the position of
the first peak moves 
from $l\sim 240$ to $l\sim 120$, and decreases in amplitude.
The higher peaks shift down in $l$ by a similar amount, with
the second peak growing to become the highest peak. 
Note that the texture curve shown here 
includes the vector and tensor contributions, which help to emphasise
the `hidden' peak at $l\sim 120$.
}
\label{fig:f3}
\end{figure}

The calculations reported here include only the $\delta_R$
contribution to $C_l$, whereas the theoretical curves shown include
all contributions, including the radiation velocity terms and the
Sachs-Wolfe integral.  In the inflationary case, the 
$\delta_R$ term alone gives a good overall representaion of
the $C_l$ spectrum for $l>100$, and particularly the peak locations.
It is therefore reasonable to expect that, at least in 
a similar situation where the Sachs-Wolfe integral and radiation
velocity terms are sub-dominant, the complete spectrum of $C_l$'s will
be similar. Computation of the Sachs-Wolfe integral, however, 
 requires 
specifying the full stress tensor $\Theta_{\mu \nu}$ of the
source, not just the combination ${\cal S}$ which was
sufficient here. In a companion paper \cite{ntnew}, I extend the definition 
of the source stress tensor to one in which the Sachs-Wolfe integral
is small, and show that 
with this extended definition of the
source, the qualitative conclusions reached here
hold for the complete $C_l$ spectrum.

Continuing with the discussion of the $\delta_R$ contribution alone, 
what general conclusions can be drawn? Any
master function $f$ can be represented by a sum of the basis 
above: since $\delta_R$ follows the same $\tau^{3\over 2}$ evolution 
up to $k\tau\sim 2.5$ for all of them, 
it follows that this will be true in the general case. Translated into
$C_l's$, this means that they
cannot have a peak
below $l\sim 120$. This is the real,
and perhaps disappointing,  causality
constraint on scaling causal sources. Can we push the first
peak to higher $l$ than in the standard
theory? Within the family considered, the limit
for the first peak is close to the adiabatic position, $l\sim 240$. 
It follows this is an upper limit 
any theory where the master functions $f(r/\tau)$
are strictly non-negative.
However, if the $f(r/\tau)$ do change sign,
the first peak may be pushed to much higher $l$.
For example adding the negative of the 
`O' case to the `X' case in Figure 2 pushes the first Doppler
peak to $l\sim 400$. Similar examples produce first Doppler peaks 
at even higher $l$.

In conclusion, I have proposed a new formalism within which
causal sources can be studied. As a first application,
I have exhibited 
a simple 
family of strictly causal sources which
produce local 
contributions to the anisotropy which
approximately 
interpolate between the standard adiabatic prediction and
the texture prediction given in \cite{ct} and \cite{durrer}. 
In a companion paper \cite{ntnew} the source model is extended to
define the full stress tensor $\Theta_{\mu \nu}$, in such a way that
the Sachs Wolfe integral is a subdominant contribution to 
the anisotropy. 
This extended model shows that
the causality constraint cannot on its own 
be used to distinguish between inflationary and non-inflationary
theories, and extra details 
are needed. 
It will be very interesting to determine
the master functions $f_\alpha$ 
and coefficients $P_\alpha$ for specific
scenarios.

I thank A. Albrecht and J. Maguiejo for useful comments.
I thank W. Hu, D. Spergel and M. White for correspondence concerning
their paper \cite{HSWnew}.
This work was partially supported by PPARC, UK, NSF contract
PHY90-21984, and the David and Lucile Packard Foundation.


\begin{thebibliography}{99}
\bibitem{proof} 
It is enough to consider the equal time correlator ($\tau=\tau'$) 
for this argument. Assume $f(r,\tau)=0$ beyond some scale $r=L$.
Consider the convolution of $f$ with itself, with argument $2L-\epsilon$.
Only the outermost nonzero shell of $f$ overlaps, and the contribution 
is proportional to $f(L-\epsilon)^2$. Since the convolution must be zero,
it follows that $f(L-\epsilon)^2$ must be zero.  Reapplying the argument for
all $L$ down to  $L=\tau$, we see that $f(r,\tau)=0$ for all $r>\tau$.
Because the full correlator (\ref{eq:c7}) 
is a sum of such 
convolutions, with non-negative coefficients, the argument shows that
{\it all} the $f_\alpha(r,\tau)$ vanish for $r>\tau$.
\bibitem{transn} 
The matter radiation transition does 
cause a  departure from exact scaling, introducing the additional
scale $\tau_{eq}$. This can be
 incorporated into
the formalism but the effect is
likely to have little impact 
on the location of the main `Doppler' peaks and I shall ignore it.
\end{thebibliography}
\end{document}